\documentclass[10pt]{article}
\usepackage{ametsoc2col}

\usepackage{amsmath}
\usepackage{textcomp}

\newcommand{\myabstract}{
The high-frequency band of the $\delta^{18}\mbox{O}$ variations in the North Greenland Ice Core Project displays fluctuation levels that increase as one approaches the onset of an interstadial (warm) period. For some of the events it is possible to establish statistical significance using Monte-Carlo simulations with a non-parametric null model with random phases and the same spectral density as the $\delta^{18}\mbox{O}$ record during the stadial periods. Similar results are found for the locally estimated Hurst exponent for the high-frequency fluctuations, and it is therefore natural to interpret these findings as so-called ``critical slowing down'' signatures, i.e. early-warning signs of tipping points. The observed ``slowing down'' is found to be similar (and perhaps even stronger) in the Younger Dryas, suggesting that there are some similarities between mechanisms of the Younger Dryas-Preboreal transition and the onsets of the Greenland interstadials. It is also verified that the temperature fluctuations during the last glacial period are characterized by long-range dependence, where the stadial periods can be approximately modeled as a $1/f$-noise. Persistent processes can take shape if the physical signal is an aggregation of several different processes, where each process responds to perturbations on a certain characteristic time scale. The results are consistent with the hypothesis that both the onsets of the Greenland interstadials and the Younger Dryas-Preboreal transition are caused by tipping points in dynamical processes with characteristic time scale of the order of decades, and that the variability of other processes on longer time scales mask the early-warning signatures in the $\delta^{18}\mbox{O}$ signal.         
}

\begin{document}

%
%
\title{\textbf{\large{Early-Warning Signals for the onsets of Greenland Interstadials\\ and the Younger Dryas-Preboreal transition}}}
%
%
\author{\textsc{Martin Rypdal}\thanks{\textit{Corresponding author address:} 
				Martin Rypdal, UiT The Arctic University of Norway, 9037 Troms{\o}, Norway
				\newline{E-mail: martin.rypdal@uit.no}}\\
\textit{\footnotesize{Department of Mathematics and Statistics, UiT The Arctic University of Norway, Norway}}
}
%
\ifthenelse{\boolean{dc}}
{
\twocolumn[
\begin{@twocolumnfalse}
\amstitle

\begin{center}
\begin{minipage}{13.0cm}
\begin{abstract}
	\myabstract
	\newline
	\begin{center}
		\rule{38mm}{0.2mm}
			\end{center}
\end{abstract}
\end{minipage}
\end{center}
\end{@twocolumnfalse}
]
}
{
\amstitle
\begin{abstract}
\myabstract
\end{abstract}
\newpage
}
\section{Introduction}
Analysis of the relative variations of the $^{18}\mbox{O}$ isotope in Greenland ice cores shows that there was a sequence of large and abrupt temperature changes during the most recent ice age. The most prominent of these changes are the transitions between the cold stadial periods and the warmer interstadial periods, during which the temperature typically increased by about 10$^\circ \mbox{C}$ within a couple of decades. The onset of the Greenland interstadials (GI) were often followed by a slow cooling, which in some cases persisted for millennia, before there were more rapid transitions back into the stadial state. These cycles are called Dansgaard-Oeschger (DO) events \citep{Dansgaard:1984eb, Dansgaard:1993fc}. In this paper we analyze the ice core record from the North Greenland Ice Core Project (NGRIP) for the time period from 60 kyr before before present (BP)\footnote{Here the present is taken to be the year Common Era (CE) 2000} to the beginning of the Holocene, and in this data set one typically recognizes seventeen DO events \citep{Svensson:2008bo}. The termination of the Younger Dryas\footnote{The last stadial period seen in the Greenland ice cores.} (YD), marks the end of the last glacial period, and this event does not define the onset of a DO cycle. However, there is little agreement in the scientific literature as to what the mechanisms for the YD were \citep{Broecker:2010ga}, and since the YD-Preboreal transition is as abrupt as the onsets of the interstadials, it is natural to include this event in this investigation.  

It is widely accepted that the onset of an interstadial period is associated with an abrupt loss of sea ice in the North Atlantic as a response to a change in the meridional overturning circulation (MOC). Positive feedback effects, such as the sea ice-albedo feedback 
and the sea ice-insulation feedback, 
can accelerate the effect of a changing ocean circulation, and cause rapid warming as a non-linear response.

\begin{figure*}[t]
\begin{center}
\includegraphics[width=15.0cm]{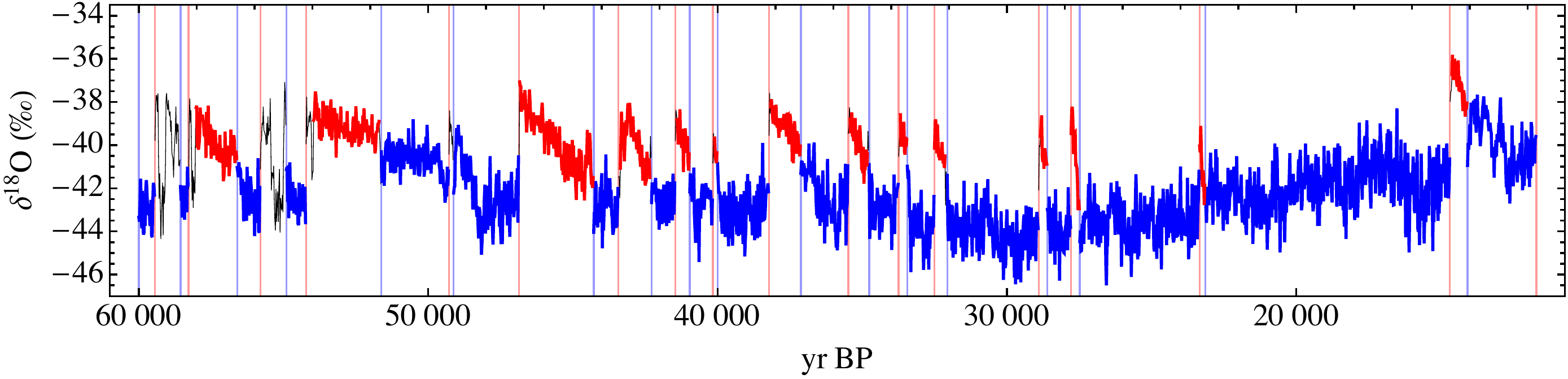} 
\caption{The NGRIP $\delta^{18}\mbox{O}$ record. The parts of the curve that are drawn in blue are defined as the cold periods, and it is these data that are analyzed for EWS. The part of the curve that is drawn in red is defined as the warm periods, and these are used to compute the PSD for the interstadial periods that is shown in Fig.~\ref{fig3}.}\label{fig1}
\end{center}
\end{figure*}

The mechanisms of the MOC variations during the last ice age, and their relation to the DO events are not well understood. It is believed that the MOC was subject to rapid changes in response to freshwater perturbations, but it is not clear which forcing agent is responsible for these changes. \cite{Grootes:1997bd}  have reported a spectral peak in the the $\delta^{18}\mbox{O}$ records from the Greenland Ice Core Project (GRIP) at a frequency corresponding to a period of about 1470 years, and it has been suggested that this periodicity is produced by the de Vries/Suess and Gleissberg solar cycles  \citep{Suess:2006dh,Sonett:1984if} (which have observed periods of 208 years and 88 years respectively). The mechanisms through which spectral peak in the $\delta^{18}\mbox{O}$ records could be linked to the shorter solar cycles, is the phenomena known as ghost resonance, and the plausibility of this explanation has been established by demonstrating that a 1470-yr periodicity in temperature can be produced from climate models if one explicitly introduce the periodicities of 208 years and 88 years in the salinity perturbations of the MOC \citep{Braun:2005fa}. Other authors have pointed out that it is difficult to establish statistical significance of  the 1470-yr periodicity in the ice-core data, and that the DO events may be triggered randomly by noise-like fluctuations in the climate system  \citep{Ditlevsen:2007jk}.

Whether the DO cycles are noise induced is of course not a close-ended question, and the answer depends to some extent on the modeling framework. Since temperature variations in general have unpredictable (or random) components on all the relevant time scales, and since the temperature fluctuation levels on Greenland during the last glacial period had a magnitude only a few times smaller than the typical temperature difference between the stadial and interstadial states, we expect that random fluctuations are important triggers of the DO events. This does not exclude the possibility that there are slow changes in the climate conditions, perhaps forced by the sun, that influence the probability of a regime shifting event.

\begin{figure}[t]
\begin{center}
\includegraphics[width=7cm]{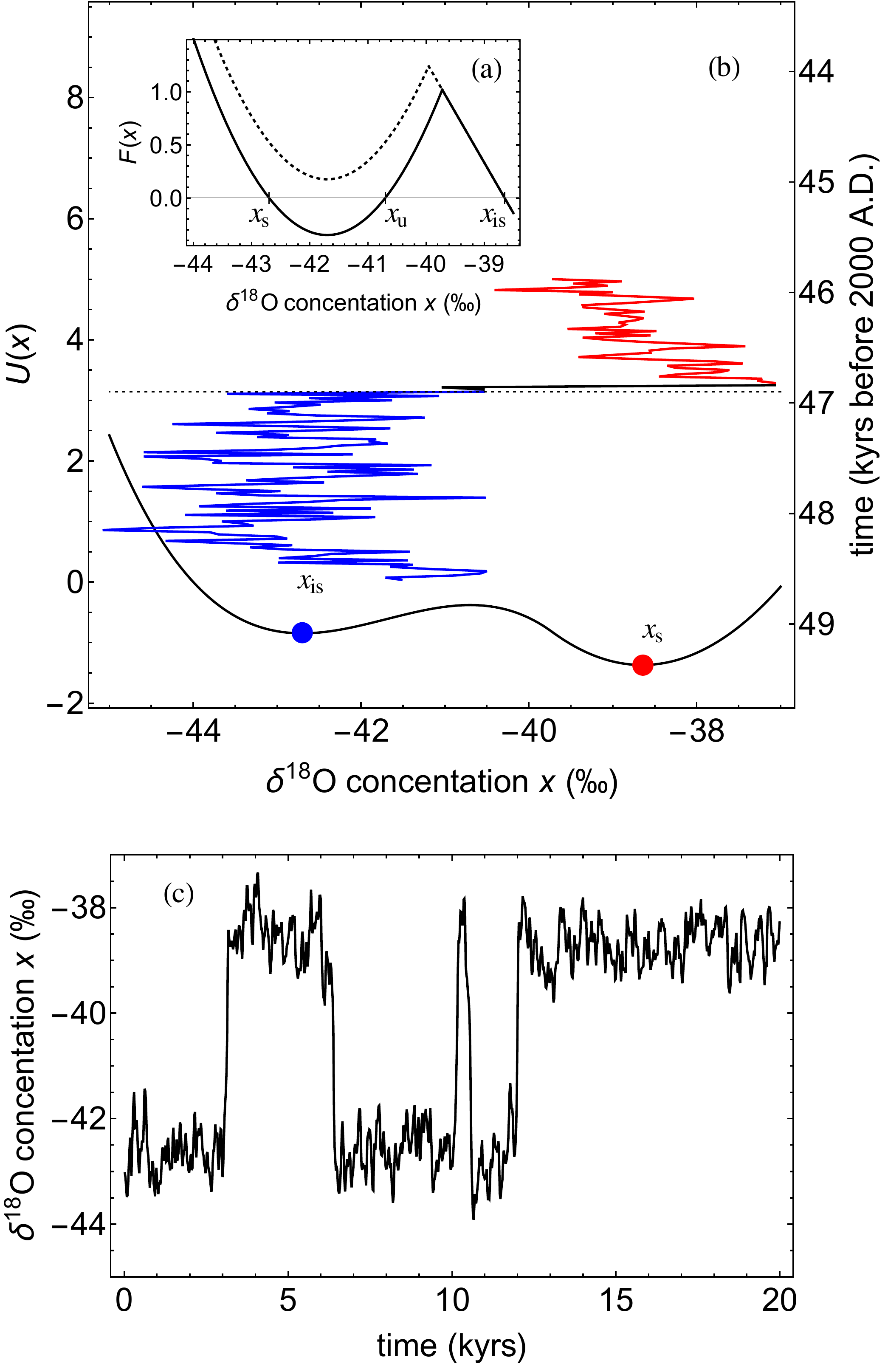} 
\caption{(a): Shows the function $F(x)$ in the example model. The dotted line shows $F(x)$ after a fold bifurcation. (b): The corresponding potential $U(x)$. The blue curve is the $\delta^{18}\mbox{O}$ signal prior to the onset of GI-12, and the red curve is the $\delta^{18}\mbox{O}$ signal during GI-12. (c): A realization of the model in Eq.~(\ref{eq1}) with $F(x)$ as shown (as the solid line) in (a).}  \label{fig2}
\end{center}
\end{figure}

Within a dynamical systems framework, these questions can be discussed in terms of the stability of the climate states, and one can use a very simple scalar model to illustrate the effect of stability weakening:
\begin{equation} \label{eq1}
dx(t) = F\big{(}x(t)\big{)}\,dt + \sigma\,dB(t).
\end{equation} 
Here $x$ can be thought of as the climate variable we seek to model, for instance the $\delta^{18}\mbox{O}$ ratio\footnote{This time series is shown in Fig.~\ref{fig1}}, $dB(t)$ is a white noise forcing of the system, and $F(x)=-U'(x)$ is a non-linear function corresponding to a potential $U(x)$. An example of a possible such model is shown in Fig.~\ref{fig2}. The system has two stable fixed points, $x_\text{s}$ and $x_{\text{is}}$, corresponding to the stadial and interstadial states. These two stable states are separated by a potential barrier with an unstable fixed point. If the noise term $\sigma dB(t)$ is sufficiently strong compared to the potential barrier, there is a non-negligible probability of a spontaneous transition between the two stable states. Such transitions are completely noise induced. 

On the other hand, we can also have a transition from the state $x_\text{s}$ to the state $x_{\text{is}}$, even in the absence of any noise, if the system goes through a bifurcation point. This means that the system depends on a slowly changing parameter $r$ in such a way that $x_\text{s}$ becomes unstable when a critical parameter value $r_c$ is reached, i.e. that $F_r'(x_\text{s}) \to 0$ as $r \to r_c$. The dotted line in Fig.~\ref{fig2}(a) shows how the stable fixed point $x_\text{s}$ is lost under a so-called fold bifurcation.  The linearization of Eq.~(\ref{eq1}) around $x_\text{s}$, 
\begin{equation}
dx(t) = - \theta \big{(}x(t)-x_\text{s})\,dt + \sigma dB(t),
\end{equation}
is known as the Langevin stochastic differential equation, and its solutions define a stochastic process called the Ornstein-Uhlenbeck (OU) process, which in discrete time is a first order auto regressive (AR(1)) process. The standard deviation of $x(t)$ in an OU process is $\sigma/\sqrt{2 \theta}$ and the autocorrelation is $e^{-\theta t}$. Since $\theta = -F_r'(x_\text{s})$, we expect increased fluctuation levels and longer correlation times in the signal $x(t)$ as the bifurcation aproaches. These signatures are called early-warning signals (EWS) of the tipping point, or ``critical slowing down'' \citep{Lenton:2012hd,Dakos:2008ga}. 

There is a key difference between a bifurcation in a completely deterministic low-dimensional dynamical system, and a tipping point in a randomly forced system, since in the latter we need not actually reach a bifurcation point in order to see a shift between two stable states. All tipping points in randomly driven systems are to some extent noise induced, and the interesting question is whether the random fluctuations are sufficient to cause a shift between the two states, or whether we can observe slow changes (perhaps forced) in the stability of the climate state. Even if EWS are not prominent features in the temperature records, observation of such structural changes may provide important insight into the mechanisms of climate tipping points.   

A few authors have already attempted to identify EWS for DO events. \cite{Ditlevsen:2010dm}  have demonstrated that it is very difficult to observe any such signatures in the Greenland ice core data, and that the ice core data are inconsistent with what we observe in typical tipping point models. On the other hand, \citet{Cimatoribus:2013fm}  have suggested to use the repeated DO events to construct an ensemble analysis that could uncover EWS that are not easily observable in the individual events. However, one must be very careful with how these ensembles are constructed. If we wish to look for EWS to the onsets of the interstadial warm periods, then the time intervals of interest are the stadial periods preceding these events. If the ensembles are constructed in such a way that the rapid cooling that marks the beginning of a stadial period is included in the ensemble members, then because of the particular timing of DO events, one is lead to the false conclusion that the fluctuation levels increase significantly as the onset of an interstadial period is approaching. If the ensemble is constructed in such a way that only stadial periods are included, then no significant EWS is seen in a standard analysis. 

It appears that these results support the findings of \cite{Ditlevsen:2010dm}, and indeed the results of this paper is evidence that the onsets of the GIs are partly noise induced and of a spontaneous nature. However, I will argue that we cannot expect to see EWS if we do not analyse individual frequency bands separately. I put forward the hypothesis that there in the stadial periods of the last glaciation, were slow changes to dynamical processes operating on decadal time scales, and that these changes are associated with a weakening of the stability of the stadial climate state on Greenland, and thereby increasing the probability of the onset of an interstadial period. In Sec.~\ref{sec2} I will show that there are anomalous fluctuation levels on the decadal time scales in the NGRIP data, and that this is an indication that one should focus on these time scales when looking for EWS. In Sec.~\ref{sec3} I show that if the NGRIP data are filtered to remove low frequency variability, then there is a slow increase in the fluctuation levels as one approaches the onsets of the interstadial periods. This result is an ensemble result where I average over the sequence of events to obtain statistically significant results. To obtain statistically significant observations of EWS in the individual events, I use wavelet analysis. The local high-frequency fluctuation levels are computed by taking the standard deviation of the wavelet coefficients corresponding to the short time scales, and the time evolution of these are analyzed. For several of the events I observe a significant increase in the standard deviations through the stadial periods. Similar results are obtained for the wavelet estimates of the local Hurst exponent, implying that the characteristic correlation length increases as the onset of an interstadial is approached. 
\begin{figure}[t]
\begin{center}
\includegraphics[width=7cm]{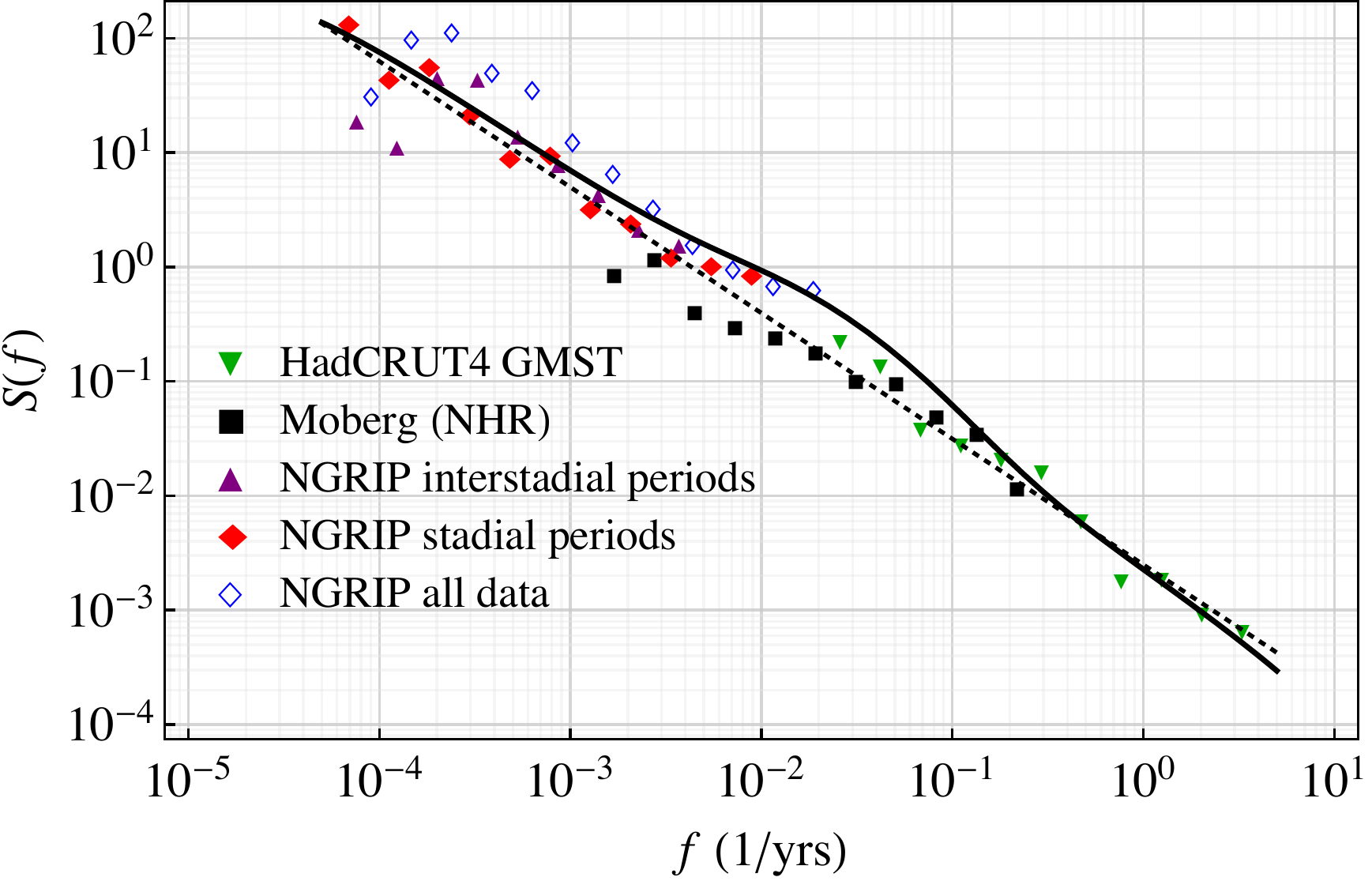}
\caption{(a): Double-logarithmic plots of the PSD $S(f)$. The analysis of the 20-yr mean NGRIP data is shown as the blue diamonds, the purple triangles and the red diamonds. The blue diamonds show the results of the analysis of the entire dataset dating back to 60 kyrs BP. The red diamonds are the results of the analysis performed on the stadial periods only, and the purple triangles are the results of the analysis of the interstadial periods only. For comparison, the green triangles represent the HadCRUT4 monthly global mean surface temperatures and the black circles is the analysis of the Moberg Northern Hemisphere temperature reconstruction. (The PSDs of the NGRIP data have been shifted to make it easier to compare with the PSDs of the two other data sets.)  The black curve is obtained from the expression in Eq.~(\ref{PSD}) (with $\beta=1.15$) by increasing the parameters $\tau_k$ corresponding to time scales between a decade and a century. } \label{fig3}
\end{center}
\end{figure}

\section{Anomalies with respect to the $1/f^\beta$ climate noise} \label{sec2}
Evidence of reduced stability on the decadal time scales during the last ice age can be observed in the estimated power spectral density (PSD) function of ice core temperature proxies. In  \citet{Rypdal:NPSTYfED}  it is shown that if the stadial and interstadial periods in the NGRIP data are analyzed separately, then fluctuations scale approximately as a $1/f$ noise, meaning that the PSD has the form $S(f) \sim f^{-\beta}$, with $\beta \approx 1$. The $1/f$ scaling observed in ice-core temperature variability is similar to what is observed in other temperature records, such as the instrumental global surface temperature and the Northern Hemisphere temperature reconstructions for the last two millennia. In fact, the $1/f^\beta$-type climate noise is what is typically observed for both global temperatures and for local temperatures\footnote{In the instrumental temperature records we find that local land temperatures scale with a lower $\beta$-exponent compared to global surface temperature and local sea-surface temperatures  \citep{Rypdal:2015iw,Lovsletten:_FzM3Wrw,Fredriksen:2016ur}. On sufficiently long time scales we expect local and global temperatures to scale with the same exponent  \citep{Rypdal:NPSTYfED} .}, and deviations from this property can be seen as anomalous. One well-known example is the El Ni{\~n}o Southern Oscillation (ENSO), which places larger fluctuation levels on the time scales of a few years than what is expected from a $1/f^\beta$ law. Another example is the large temperature variability on the decadal time scales observed in the Greenland ice cores. In Fig.~\ref{fig3} I plot the estimated PSD of the 20-yr average $\delta^{18}\mbox{O}$ variations in the NGRIP ice core. The blue diamonds is the periodogram estimation for the entire time series, whereas the red diamonds and the purple triangles are estimated using only the stadial and interstadial periods separately. This can be done using the Lomb-Scargle periodogram \citep{Lomb:1976bo}, which is an estimation technique for the PSD which does not require the signal to be sampled at equal time intervals. As we see from the figure, the PSD deviates from the $1/f^\beta$ law for frequencies corresponding to time scales shorter than a few centuries. As I will explain in the following, this effect can be taken as an indication that the processes that dominate the temperature signal on these time scales have weaker stability than what is predicted from a $1/f^\beta$ assumption. The argument behind this claim is that the scaling of the climate noise is a reflection of the fact that the climate system consists of many components that respond to perturbations on different time scales, and it is difficult to identify any characteristic time scales in the temperature records. As a straw-man model, we can think of the temperature signal as an aggregation of processes 
\begin{equation*}
T(t) = \sum_k T_k(t),
\end{equation*} 
where each term $T_k(t)$ is a (possibly) non-linear and stochastic description of the temperature variations at the time scale $\tau_k$. As linearized descriptions of the components $T_k(t)$ we can write the stochastic differential equations 
\begin{equation*}
dT_k(t) + \theta_k T_k(t)\,dt = c_k \,dB_k(t) \text{~ with ~} \theta_k = \frac{1}{\tau_k},
\end{equation*} 
and from this (assuming independence if the noise processes $dB_k$) the PSD of the aggregated signal $T(t)$ becomes 
\begin{equation} \label{PSD}
S(f) = \sum_k \frac{|c_k|^2}{\tau_k^{-2}+(2 \pi f)^2}.
\end{equation} 
The aggregated process $T(t)$ can be made to approximate a $1/f^\beta$ noise of we choose the time scales $\tau_k$ to be exponentially spaced, i.e. $\tau_k = a^k \tau_0$ for some parameter $a>0$. Here $\tau_0$ is some reference time scale, for instance $\tau_0 = 1$ yr. In addition we need to require the that 
$|c_k|^2 = a^{2-\beta} |c_{k+1}|^2$. If this is the case we have the approximate relation $S(af) \approx a^{-\beta} S(f)$, so if $\beta \approx 1$ the signal $T(t)$ will be consistent with the scaling observed in ice-core temperature records. 

Using this simple straw-man description we can now explore the effect of reducing the stability of some of the components $T_k$. For instance, if one of the components $T_k(t)$ is well-described by a non-linear model that approaches a ``tipping point'', then in the linearized model we will see $\theta_k \to 0$, corresponding to strong increase in the characteristic time scale $\tau_k$. This will lead to a deviation from the (approximate) $1/f^\beta$ law. In fact, this effect is completely consistent with our observations for the NGRIP data. The black curve in Fig.~\ref{fig3} is obtained from the expression in Eq.~(\ref{PSD}) (with $\beta=1.15$) by increasing the parameters $\tau_k$ corresponding to time scales between a decade and a century (using $a=2$ and $\tau_0=1$ yr). The effect is a ``flattening'' of the PSD on time scales shorter than a few centuries, similar to what is estimated in the NGRIP data. 

The observation discussed above is an indication that the high-frequency fluctuations in the NGRIP data are of interest when searching for EWS, but in it self this observation does not present any EWS, since it does not uncover any temporal changes in the stability of the stadial climate state. Such changes will be discussed in the following section.

\begin{figure*}[t]
\begin{center}
\includegraphics[width=15.0cm]{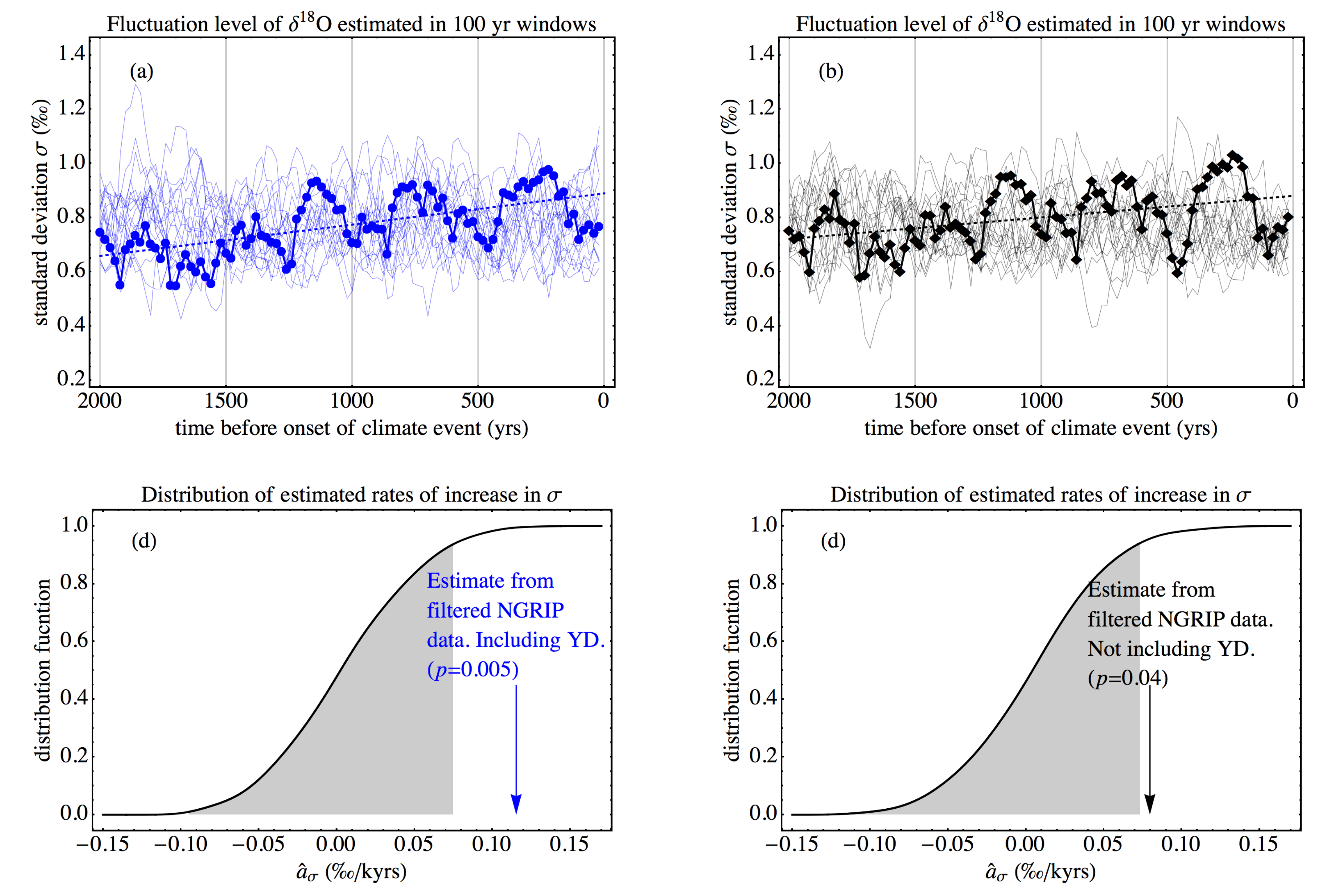} 
\caption{(a): The fluctuation level in 100-yr windows of the filtered $\delta^{18}\mbox{O}$ signal as a function of the time before the sudden onset of the warm period. The dotted line is a linear fit $\hat{a}_\sigma = 0.11 \text{~\textperthousand}/\mbox{kyrs}$. The thin curves are the corresponding fluctuation levels in a null-model which is constructed by taking the PSD of each cold period and randomizing the phases. (b): As in (a), but in this case the YD is not included as one of the cold periods. The dotted line has the slope $\hat{a}_\sigma = 0.07 \text{~\textperthousand}/\mbox{kyrs}$. (c): The distribution function of the linear fits $\hat{a}_\sigma$ under the null model. The shaded area represents the 95\% confidence of $\hat{a}_\sigma$ under the null model and the arrow marks the observation $\hat{a}_\sigma = 0.11 \text{~\textperthousand}/\mbox{kyrs}$. (d): As in (c), but in this case for the analysis that does not include the YD. The arrow marks the estimate $\hat{a}_\sigma = 0.08 \text{~\textperthousand}/\mbox{kyrs}$.} \label{fig4}
\end{center}
\end{figure*}

\begin{figure*}[t]
\begin{center}
\includegraphics[width=15.0cm]{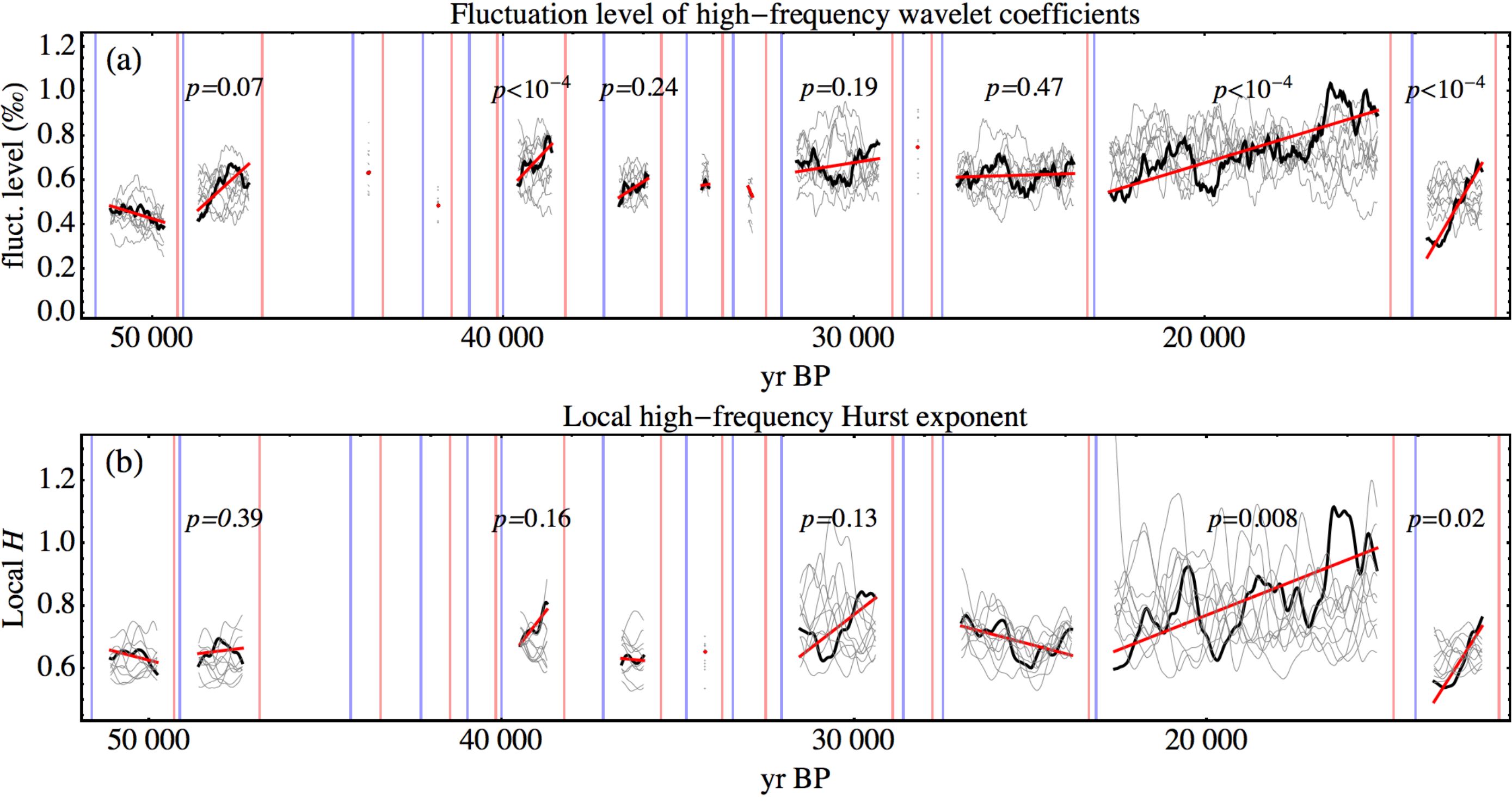} 
\caption{(a): Shows the wavelet fluctuation level $\sigma(t)$ defined by Eq.~\ref{wavelet}. The red curves are linear fits to $\sigma(t)$ in each cold period, and the $p$-values are obtained by estimating the distribution function for the linear slopes using a Monte Carlo simulation (with the null model that is constructed by randomizing the phases). (b): As in (a), but for the locally estimated Hurst exponent.} \label{fig5}
\end{center}
\end{figure*}

\section{Analysis and results} \label{sec3}
If one attempts to model the NGRIP $\delta^{18}\mbox{O}$ times series as a single randomly forced scalar dynamical system with two stable states, then any parameter choice that corresponds to realistic fluctuation levels in the the stadial and interstadial states will lead to spontaneous ``jumps'' between the two states. This is a simple consequence of the ratios between the fluctuation level and the temperature difference between the stadial and interstadial states. This is illustrated in Fig.~\ref{fig2} where I show an example of such a model, where the parameters are chosen so that the OU models (that are obtained by linearization around the stadial and interstadial sates) have standard deviations equal to the sample standard deviations of the stadial and interstadaial periods in the NGRIP time series. The fixed points are chosen according to the averages of $\delta^{18}\mbox{O}$ in the stadial and interstadial periods before and after the onset GI-12.\footnote{We refer to \citep{Svensson:2008bo} for the enumeration of the GIs.} Fig.~\ref{fig2}(c) shows a realization of this model with fixed parameters, and we see that there are transitions between the two states even in the absence of any slowly varying parameter changes, i.e,. completely noise induced shifts. 

However, as I discussed in Sec.~\ref{sec2}, it is reasonable to model the $\delta^{18}\mbox{O}$ signal as an aggregation of signals, where some of the terms do not experience any regime shifts, but nevertheless contribute to the fluctuation level. It is then possible that the shifts {\em do} require reduced stability of the stadial climate state. If this is the case, we should in principle observe EWS, but these may be be masked by the variability of the other terms contributing to the aggregated signal. A natural approach for uncovering EWS is then to filter the NGRIP data and analyze certain frequency bands. As also discussed in Sec.~\ref{sec2}, I have indications that the dynamical processes associated with reduced stability have characteristic time scales shorter than a century, and therefore I will analyse the high-frequency band of the NGRIP data. 

The first step in this analysis is to identify stadial periods and the onset times for the interstadial periods. In total I analyze eighteen climate events. These include the onsets of GI 1-17 as well as the YD/Preboreal transition. I have used the onset dates for the interstadial periods (and the date for the YD-Preboreal transition) as given by \cite{Svensson:2008bo}. These dates determine the end of the cold periods which are investigated for EWS. The start dates for the cold periods are chosen such that they do not include the very sudden temperature declines that often occur in the DO cycles. These sudden temperature changes (which are believed to be linked to slowdowns of the thermohaline circulation coupled with sea-ice formation) can themselves be seen as tipping points \citep{Lenton:2012hd}, and should not be viewed as a part of the destabilization of the cold state. The cold periods we have chosen to analyze are drawn as blue curves in Fig.~\ref{fig1}.  

In Fig.~\ref{fig4} I show the results of an analysis where I consider each cold period as an ensemble member. The $\delta^{18}\mbox{O}$ time series is filtered by subtracting a 100-yr moving average, and for the filtered signal I have computed the standard deviation in running 100-yr windows. For the cold periods (those drawn in blue in Fig.~\ref{fig1}), the results are organized by averaging the standard deviation over all 100-yr time windows that precede the onset of an interstadial period by a certain number of years. In this way I obtain an ensemble estimate of the fluctuation level in the $\delta^{18}\mbox{O}$ signal as a function of the time before the sudden onset of the warm period. In Fig.~\ref{fig4}(b) I have plotted the fluctuation level when I have included all cold periods with duration longer than 300 yrs, but not including the YD. The dotted line is a linear fit with a slope $\hat{a}_\sigma=0.08 \text{~\textperthousand}/\mbox{kyrs}$. 

This increasing slope is significantly larger than zero, with a $p$-value of $0.04$. The significance is tested by constructing signals that have the same PSD as the cold-period signals, but where the phases are randomized. For each of the cold periods I compute the discrete Fourier transform (DFT) of the $\delta^{18}\mbox{O}$ signal, and for each frequency the square root of its modulus is multiplied by a factor $e^{i \phi}$, where $\phi$ is a random angle chosen with respect to the uniform distribution on the interval $[0,2\pi)$. The inverse DFT is applied to the resulting time series, before taking the real parts and adjusting the standard deviations by a factor $\sqrt{2}$.\footnote{Since we disregard the imaginary part of the constructed signal, this adjustment is needed in order for the synthetic signals to have the same standard deviations as the cold periods in the  $\delta^{18}\mbox{O}$ signal.} The thin curves in Fig.~\ref{fig4}(b) show how the standard deviations in 100 yr windows of the (filtered) synthetic realizations depend on the time before the onset of the interstadial periods. In a large ensemble of realizations the pseudo-slopes $\hat{a}_\sigma$ are computed, and the distribution function $P(\hat{a}_\sigma)$ of these is obtained using a smooth kernel estimator \citep{Rosenblatt:1956jr}. The estimated distribution function is shown in Fig.~\ref{fig4}(d). The arrow in this figure shows the value $\hat{a}_\sigma=0.08 \text{~\textperthousand}/\mbox{kyrs}$ estimated from the $\delta^{18}\mbox{O}$ signal, and the gray area under the curve marks the 95\% confidence interval for $\hat{a}_\sigma$ under the null model. The $p$-value is computed as $p=1-P(0.08 \text{~\textperthousand}/\mbox{kyrs})$. In Fig.~\ref{fig4}(a) and Fig.~\ref{fig4}(c) I show the results of the same analysis, but in this case the YD-Preboreal transition is included in the analysis, which in practice means that the YD is included as one of the cold periods under investigation. When the YD is included the estimate becomes $\hat{a}_\sigma = 0.11 \text{~\textperthousand}/\mbox{kyrs}$, whereas the distribution $P(\hat{a}_\sigma)$ changes very little, and the statistical significance is improved to $p=0.005$. 

The ensemble results presented above show that there is a tendency for the fluctuation levels to increase towards the sudden termination of the Greenland stadials. However, it does not tell us whether these EWS are observable in the individual climate events. In order to analyze the individual events I use the wavelet transform 
\begin{equation*}
W(t,\tau) = \frac{1}{\sqrt{\tau}} \int x(t') \psi\Big{(} \frac{t-t'}{\tau} \Big{)} dt'
\end{equation*}
and estimate the time-varying fluctuation levels in the high-frequency band by averaging over the time scales $0<\tau<\tau_c$ and over time windows of length $\Delta t$:  
\begin{equation} \label{wavelet}
\sigma^2(t) = \frac{1}{\Delta t} \int_0^{\tau_c} \int_{t-\Delta t/2}^{t+\Delta t/2} |W(t',\tau)|^2  dt' d\tau. 
\end{equation}
I have used $\tau_c=50$ yrs and $\Delta t=200$ yrs, and the time variation of $\sigma(t)$ for each cold period is shown in Fig.~\ref{fig5}(a). Linear fits to $\sigma(t)$ in each cold period are drawn in red, and realizations of $\sigma(t)$ for the synthetic signals (using the same null model as described above) are plotted as the thin curves. The distribution function for the linear pseudo trends in the null model is obtained via a smooth kernel estimator, and using this I compute $p$-values for the linear increases in $\sigma(t)$. These $p$-values are shown in the figure. We have $p<10^{-4}$ prior to the onset of the YD-Preboreal transition and prior to the onsets of GI-1 and GI-8. Prior to GI-12 we have significance at the $0.1$-level, and in a majority of the cold periods we have increasing trends in $\sigma(t)$.

In Fig.~\ref{fig4}(b) we show the time dependence of the locally estimated Hurst exponent $H$. This is estimated via the relation 
\begin{equation*}
\langle |W(t,\Delta t)|^2 \rangle \sim \Delta t^{2H-1}, 
\end{equation*}
i.e. a linear fit is made to $\log |W(t,\Delta t)|^2$ as a function of $\log \Delta t$. The fluctuations $\langle |W(t,\Delta t)|^2 \rangle$ are estimated in 200 yr windows and only the time scales shorter than $60$ yrs are used. Since only the high frequency fluctuations are used to estimate $H$ it is more appropriate to think of it as a local smoothness exponent than as a scaling exponent\footnote{If we interpret $H$ as a scaling exponent then it is related to $\beta$ via the relation $\beta=2H-1$.}. Nevertheless, a time varying Hurst exponent estimate that increases in time is consistent with an increase in correlation length in the high-frequency band, and it is thus expected in association with stability loss. As with the high-frequency wavelet fluctuation level, we see strongly significant increases in $H$ before the onsets of GI-1 and the YD-Preboreal transition approaches. Strong increases are also seen before GI-8 and GI-4. 

I have chosen to use a Paul mother wavelet when estimating the fluctuation level $\sigma(t)$ and to use a Morlet wavelet when estimating $H$. The reason for these choices is that I want to optimize the time resolution in the $\sigma(t)$ estimation and the scale resolution in the estimation of $H$  \citep{DeMoortel:gs}.

\section{Discussion and concluding remarks}
This paper presents both new results and new methods. The new methods include combining high-pass filtering with the ensemble construction presented by \cite{Cimatoribus:2013fm}, as well as using the wavelet transform to discern time-varying fluctuations in the high-frequency band. Another important aspect is the statistical significance testing, which is based on a non-parametric null model with random phases. Due to the ``flattening'' of the PSD at high frequencies, the application of a parametric model such as a fractional Gaussian noise (fGn) will lead to a misrepresentation of the fluctuation levels either on the short time scales or on the long time scales (depending on which time scales are emphasized in the parameter estimation). In either case it will provide an inaccurate model for the distribution of pseudo-trends in the local fluctuation levels. For instance, if I were to apply an fGn null-model using standard parameter estimation methods, then this model would underestimate the high-frequency fluctuation levels, and as a consequence I would obtain much lower $p$-values for the EWS.      

The methods described above are different from the those used by \cite{Lenton:2012hd} and \cite{Dakos:2008ga}, who focus on the lag-1 autocorrelation and the Hurst exponent estimated using de-trended fluctuation analysis (DFA). While these approaches are very robust, they have some disadvantages. A problem with the lag-1 autocorrelation is its sensitivity to trends, and to low frequency variability, and the DFA estimator is known to resolve time scales poorly. I also note that the filtering applied in \cite{Lenton:2012hd} and \cite{Dakos:2008ga} is meant as a de-trending, and care is made as to not filter out the low-frequency variability in the signal, while in this paper I {\em do} wish to remove the slow fluctuations that are masking the EWS.

The EWS we find for the YD-Preboreal transition are consistent with results of \cite{Lenton:2012hd} and \cite{Dakos:2008ga}. We also find strong EWS for the onset of GI-1 (the so-called Bølling-Aller{\o}d warming) and GI-8, and seen as an ensemble, we find significant EWS for the onsets of the interstadial periods. The results show that there are dynamical structures related to the DO cycles that experience reduced stability prior to the onset of a sudden warming. On the other hand, these signals are not prominent features in the temperature signal and I agree with \cite{Ditlevsen:2009gq}  that the onsets of GIs must to some extent be seen as random and unpredictable events. 

The observation that the stadial climate on Greenland experience reduced stability prior to the onsets of the interstadials is complementary to the findings of \cite{Livina:2010bw}, who have made similar observations (using very different methods) for the interstadial climate states. The study of \cite{Livina:2010bw} is consistent with the observation of EWS in climate models forced through a shut down of the Atlantic Thermohaline circulation \citep{Lenton:2012hd}.

 \acknowledgment  This work has received support from the Norwegian Research Council under contract 229754/E10. The author thanks K. Rypdal, H.-B. Fredriksen, T. Nilsen and O. L{\o}vlsetten for useful discussions and suggestions.

\end{document}